\documentclass{aa}

\usepackage{graphicx}
\usepackage{longtable}
\usepackage{graphicx,wasysym}
\usepackage{longtable}
\usepackage{natbib}
\bibpunct{(}{)}{;}{a}{}{,}
\usepackage{color}
\usepackage{amssymb}

\def\Msol{M_\odot}
\def\gr{$\gamma$-ray\ }
\def\grs{$\gamma$-rays\ }

\def\alphan{{\alpha_\mathrm{n}}}

\begin{document}

\title{On the potential of the Cherenkov Telescope Array for the study of 
cosmic-ray diffusion in molecular clouds}
\author{
 G.~Pedaletti \inst{1},
 D. F. Torres \inst{1,2},
 S. Gabici \inst{3},
 E. de O\~{n}a Wilhelmi \inst{4},
 D. Mazin \inst{5, 6},
 \and V. Stamatescu \inst{5}
}

\institute{
Institut de Ci\`encies de l'Espai (IEEC-CSIC),
              Campus UAB,  Torre C5, 2a planta,
              08193 Barcelona, Spain
\and 
Instituci\'o Catalana de Recerca i Estudis Avan\c{c}ats (ICREA)
\and
Astroparticule et Cosmologie (APC), CNRS, Universit\'{e} Paris 7 Denis Diderot, Paris, France
\and
Max-Planck-Institut rf\"{u}r Kernphysik (MPIK), P.O. Box 103980, 69029 Heidelberg, Germany
\and 
IFAE, Edifici Cn., Campus UAB, E-08193 Bellaterra, Spain
\and
now at: Max-Planck-Institut f\"ur Physik, D-80805 M\"unchen, Germany
}
\date{Received: 17 October 2012 / Accepted: 22 December 2012}

\abstract{}{Molecular clouds act as primary targets for cosmic-ray interactions and are expected to shine in \grs as a by-product of these interactions. Indeed several detected \gr sources both in HE and VHE \grs (HE: 100 MeV $<$ E $<$ 100 GeV; VHE: E $>$ 100 GeV) have been directly or indirectly associated with molecular clouds. Information on the local diffusion coefficient and the cosmic-ray population can be inferred from the observed \gr signals.
In this work we explore the capability of the forthcoming Cherenkov Telescope Array Observatory (CTA) to provide such measurements.}
{We investigate the expected emission from clouds hosting an accelerator, surveying the parameter space for different modes of acceleration, age of the source, cloud density profile, and cosmic-ray diffusion coefficient. }
{We present some of the most interesting cases for CTA regarding this science topic. The simulated \gr fluxes depend strongly on the input parameters. 
In several cases, we find that it will be
possible to constrain both the properties of the accelerator and the
propagation mode of cosmic rays in the cloud from CTA data alone.}{}


\authorrunning{G. Pedaletti et al}
\titlerunning{CTA and cosmic ray diffusion in molecular clouds}
\maketitle

\section{Introduction}
Emission in HE-VHE \grs is expected in spatial
coincidence with molecular clouds, resulting from the hadronic
interaction between cosmic-ray (CR) particles and the dense material
in the cloud acting as a target. Indeed, some MCs have been detected
in \grs in both the GeV and TeV domain \citep[see, e.g., ][]{gc_hess,ic443_magic,w28_hess,w44_agile,mc_fermi,aha_cta_review}. Moreover, it has been suggested that some of the as yet unidentified \gr sources might also be MCs illuminated by CRs that escaped from an accelerator located inside the cloud or in its proximity \citep{montmerle79,aha_ato,gabici2007,rodriguez2008}. In such cases the modeling of the emission involves the parametrization of the diffusion of charged particles. The diffusion coefficient is in general considered to be energy-dependent \citep{ginzburg1964}, with lower energy particles diffusing more slowly than higher energy ones, under the same medium conditions. 
When those diffused charged particles (protons or heavier nuclei) interact with target material in a density enhancement of the interstellar medium, such as a molecular cloud located in the vicinity of the accelerator, significant \gr emission is expected due to the production and subsequent decay of neutral pions. \gr emission produced in massive molecular clouds was predicted long ago \citep[see e.g.][]{black_fazio,morfill1984,aha_passive}. 
The study of this emission is extremely useful in unveiling the physics of CR sources. 
Due to energy dependent propagation effects the \gr spectrum from the molecular clouds may differ significantly from the spectrum observed at (or closer to) the accelerator. This may explain discrepancies in the particle spectral indeces inferred from the same source at different frequencies, even if all particles, leptons and hadrons, are accelerated to the same power-law at the source.  
In this scenario, a great variety of \gr spectra is expected, depending on several parameters, including: the age of the acceleration, the distance between the cloud and the accelerator, the duration of the injection of CRs and their diffusion coefficient. This can produce a variety of different GeV--TeV connections, some of which could explain the observed phenomenology \citep[see, e.g.,][]{funk2008,tam_gevtev}. 
Injection and propagation of CR have been recently put forward to explain a number of the TeV sources currently known, especially those for which there is no, or there is a spatially displaced, GeV counterpart \citep[e.g. see][]{fujita2009,liandchen2010,gabici_w28,torres2010,ohira2011}.

Given the expected CTA angular resolution and sensitivity, variations in flux at less than 5 pc bins at 5kpc distances could be resolved. Changes in the CR spectrum could be derived accordingly, leading even to, e.g., the derivation of a diffusion coefficient as a function of energy and position ($D(E,r)$). The measurement of such spatial variability in the diffusion coefficient would be an important result in CR physics. 

In this paper, the \gr emission due to an accelerator inside a molecular cloud is calculated. The expected CTA measurement of such emission is then derived taking into account the simulated CTA response functions. The CTA Observatory is described in Section 2. The calculation of \gr emission and the physical parameters of the scenario are described in Section 3. The simplified case of flat density of the target material (i.e. flat density profile of the molecular cloud) is investigated in Section 4 along with the CTA capabilities in distinguishing the parameter space. Small and nearby clouds are investigated in Section 5. Section 6 deals qualitatively with a more realistic case of a peaked density profile. Conclusions are given in Section 7.
\section{The Cherenkov Telescope Array (CTA)}
CTA is an international project for the development of the next generation ground-based \gr instrument \citep[see][]{dc2010}. The detection of \grs ($E>$10 GeV) with ground-based facilities is possible thanks to the imaging atmospheric Cherenkov technique \citep{book_weekes}. VHE \grs interact with nuclei in the atmosphere producing a cascade of particles, where velocities are larger than the speed of light in the medium, leading to Cherenkov light emission. The resulting Cherenkov light flashes may be imaged by Imaging Atmospheric Cherenkov Telescopes (IACTs) \citep[for a recent review, see ][]{vhereview2009}. The shower images are then used to reconstruct the energy and direction of the original particle. Particle cascades can also be initiated by CRs and constitute the main source of background. In this case the showers are generally broader than those initiated by primary photons. Gamma-hadron separation can be achieved thanks to the differences in size and shape of the shower images. The showers illuminate a pool at the ground level. The radius of the light pool depends on the height of the observatory and on the energy of the primary photon. An array of IACTs allows for better sampling of the Cherenkov light distribution of a given event. From a stereoscopic view of the same event, the reconstruction of the direction of the primary photon and the background rejection are improved with respect to a stand-alone telescope observation. 

CTA will significantly advance on the present generation
IACTs: it will feature an order of magnitude improvement in
sensitivity at the core energy range of 1 TeV, improve in its angular
and energy resolution, and provide wider energy coverage, see \citet{dc2010}. Indeed, the array is expected to have an unprecedented sensitivity down to $\sim$50 GeV and above $\sim$50 TeV, establishing a strong link to the satellite-based operations at low energies, namely the Large Area Telescope on board the {\it Fermi} satellite, see \citet{atwood2009} and water Cherenkov experiments at the highest energies \citep[e.g, HAWC, see][]{hawc2010}. The gain in sensitivity is due to the increase in the number of telescopes. The widening of the explored energy range is due to a combination of different-sized telescopes in different parts of the light pool. 
Large size telescopes (LST, with a dish of $\sim$23 m) will be placed at the center of the array. Thanks to their large mirror area, dim flashes from the low energy events ($\sim$50 GeV) are expected to be reconstructed. Tens of medium size telescopes (MST, with a dish of $\sim$11 m) will be placed in a surrounding ring, covering a large fraction of the light pool and thus enhancing the reconstruction of medium energy ($\sim$1 TeV) events. Finally, the outer regions will be composed of small size telescopes (SST, with a dish of $\sim$7m) enlarging the effective area of the array for the bright but rare high energy events (above $\sim$50 TeV). Both a southern and a northern hemisphere observatory are foreseen. 
\section{An accelerator inside a cloud}\label{sec:accelerator}
If a power-law energy spectrum ($J_\mathrm{p}(E_\mathrm{p})=K E_\mathrm{p}^{-\gamma}$) is assumed for the intensity of primary CRs, the resulting \gr spectrum due to hadronic interactions would also follow a power-law spectrum ($F(E) \propto E^{-\Gamma}$). However, if we consider an energy-dependent diffusion coefficient, the CR spectrum may differ from a simple power-law near the acceleration site. The spectrum of the accelerated CR can be expressed as $J_\mathrm{p}(E_\mathrm{p},r,t)=(c/4\pi)f$, where $f(E_\mathrm{p},r,t)$ is the distribution function of protons at time $t$ and distance $r$ from the source. The distribution function satisfies the diffusion-loss equation \citep[e.g.,][]{ginzburg1964}
\begin{equation}
\frac{\partial f}{\partial t}=\frac{D(E_\mathrm{p})}{r^2}\frac{\partial}{\partial r}r^2\frac{\partial f}{\partial r}+\frac{\partial}{\partial E_\mathrm{p}}(Pf)+Q,
\end{equation}
where $P=-dE_\mathrm{p}/dt$ is the continuous energy loss rate of the particles, $Q=Q(E_\mathrm{p},r,t)\delta(R)$ is the source function (for injection), and $D(E_\mathrm{p})$ is the diffusion coefficient. Here, we assumed that the source is point-like and located at the origin of the coordinate system. Solutions to this equation have been extensively studied for different cases, considering either spatially constant diffusion coefficient \citep{atoyan1995,aha_ato,rodriguez2008,gabici2009} or no CR accelerator near the cloud, i.e. Q=0, \citep[passive clouds where the only \gr emission arises from the contribution of the CR background,][]{gabici2007}. 
We investigate the case of an accelerator positioned at the center of a
molecular cloud.
This is an idealized case, but it allows to study the impact of an
enhancement of CR content, above and beyond the passive cloud case. The
\gr emission can be calculated in concentric shells of increasing radius,
each shell retaining the footprint of the diffusion coefficient and of the
cloud density. The study of such footprint is done for a simple
symmetrical and homogeneous system, where expected spectral and
morphological behaviors can be shown clearly.

The acceleration and diffusion processes are computed following the approach of \cite{aha_ato}. The diffusion coefficient is assumed to depend on the CR energy only, as: $D(E_\mathrm{p})=D_\mathrm{10} (E_\mathrm{p}\, / \, 10\,\textrm{GeV})^{\delta}$ cm$^{2}$ s$^{-1}$.
More details of the flux calculation are given in the Appendix \ref{app:grid}. 
The resulting flux in \grs is mainly dependent on the diffusion coefficient, the age of the accelerator, the type and spectrum of injection of accelerated particles, and on the density and mass of the cloud. An impulsive source of particles corresponds to the case when the bulk of relativistic cosmic-rays are released during times much smaller than the age of the accelerator itself. When the timescales are comparable, the source is referred to as a continuous injector. 
All the parameters are free and might assume slightly different values to those studied here. The intervals for the values of the parameters adopted in this work are given below:
\begin{itemize}
 \item Diffusion coefficient: Slow to fast (e.g., $D_{10}=[10^{26} .. 10^{28}]$ cm$^{2}$ s$^{-1}$);
\item Diffusion coefficient energy dependence: $\delta= [0.3 .. 0.6]$;
\item  Age of the accelerator: $[10^{3}$ .. $10^{5}]$ years
\item  Type of accelerator: Impulsive / Continuous
\item  Spectrum of injection: $\gamma$= $[2.0 .. 2.5]$;
\item  Fraction of energy in input (total energy in form of CRs $\mathrm{W_p}=\eta10^{50}$ erg): $\eta=[0.3 .. 3]$. 
\end{itemize}
However, we mainly concentrate, as an example, on the case were the injection slope is $\gamma=2.2$ and the diffusion coefficient dependence on energy is $\delta=0.5$. This pair of parameters satisfies the observed CR spectrum data. Indeed, with these values, the index of the equilibrium spectrum in the galaxy is expected to be $\gamma+\delta=2.7$. The typical value of the diffusion coefficient in the galaxy is $D_{10}=10^{28}$ cm$^{2}$ s$^{-1}$ \citep{ber_book}. However, this value is very uncertain and depends on the level
of the magnetic turbulence in which particles propagate. For example, higher level of turbulence will in general lead to a suppression of diffusion. The total energy input is taken as $\mathrm{W_p}=10^{50}$ erg ($\eta=1$) in the impulsive case, while the energy injection rate in the continuous case is of $L_\mathrm{p}=10^{37}$ erg s$^{-1}$, resulting in the same total input for accelerators with an age of a few hundreds of thousand years.

The spectra of \grs produced by proton-proton interactions have been computed
following \cite{aha_ato}, where a delta-function approximation has been
used to model the interaction cross section. 

\subsection{CR sea penetration in the cloud}

\begin{figure}[!h]
\begin{center}
\includegraphics[width=0.98\linewidth]{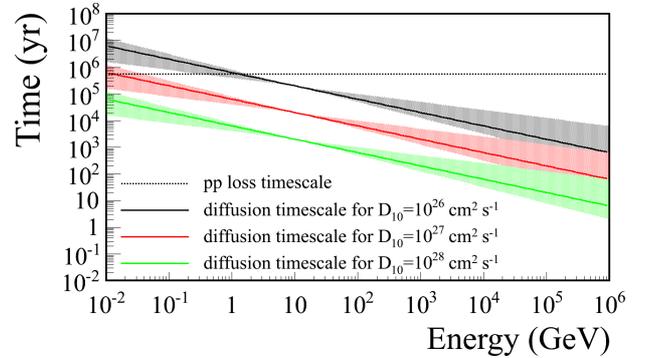}
\caption{Comparison of relevant timescales calculated via Eq. (\ref{eq:taupp}) and Eq. (\ref{eq:taudiff}). The dashed line represents the loss timescale through pp channel. The shaded areas represent the diffusion timescale for the interval $\delta= [0.3 .. 0.6]$. The solid lines are for $\delta= 0.5$. The crossing of the dashed and solid lines represent the minimal energy at which total cloud penetration is fulfilled for a diffusion coefficient of $D_\mathrm{10}=10^{26}$ cm$^{2}$ s$^{-1}$ (black), $D_\mathrm{10}=10^{27}$ cm$^{2}$ s$^{-1}$ (red), $D_\mathrm{10}=10^{28}$ cm$^{2}$ s$^{-1}$ (green).}
 \label{fig:penetration}
\end{center}
\end{figure}

The CR background (also referred to as `sea') 
is ubiquitous in the Galactic plane. For the energy range considered here, it is assumed to be well described by the locally measured differential spectrum as \citep[e.g., see ][]{pamela_cr}
\begin{equation}
 J_\odot(E_p)\simeq 1.5 E_\mathrm{p,GeV}^{-2.7} \mathrm{cm}^{-2} \mathrm{s}^{-1} \mathrm{sr}^{-1} \mathrm{GeV}^{-1}.
\end{equation}
Therefore, to compute the total CR content inside the cloud, it is necessary to calculate the degree of penetration of the CR sea in the simulated cloud. The comparison of relevant timescales (pp losses and diffusion timescale) has been done following, e.g., \cite{gabici2007}.
The energy loss timescale for proton interactions is 
\begin{equation}\label{eq:taupp}
 \tau_{pp}=\frac{1}{n_\mathrm{H} c \kappa \sigma_{pp}},
\end{equation}
where $n_\mathrm{H}$ is the density of the gas, $c$ is the speed of light, $\kappa$ is the inelasticity (assumed to be $\kappa\sim$0.45 throughout the paper) and $\sigma_{pp}=33$ mb is the cross-section of the process. $\sigma_{pp}$ is mildly dependent on the energy of the particles, however, for the energies considered here, the assumption of energy independence is satisfactory \citep{aha_ato}.
The diffusion timescale, i.e. the time it takes a CR to diffuse
from the edge to the centre of the cloud, can be expressed as 
\begin{equation}\label{eq:taudiff}
 \tau_{diff}=\frac{R_\mathrm{cloud}^2}{6 D(E_\mathrm{p})},
\end{equation}
where $R_\mathrm{cloud}=20$pc is the radius of the cloud in the example investigated here. For a mass of $10^5 M_\odot$, this results in a uniform density of $n_\mathrm{H}=130$ cm$^{-3}$, rather typical for a molecular cloud. The timescale of Eq. \ref{eq:taudiff}, in the impulsive case, represents the time at which the maximum of particle flux is reached at the distance $R_\mathrm{cloud}$ \citep{aha_ato}.
Penetration is ensured for particles of energies higher than the energy resulting from equating Eq. (\ref{eq:taupp}) and Eq. (\ref{eq:taudiff}). For the smaller value of the diffusion coefficient adopted here (i.e. $D_{10}=10^{26}$ cm$^{2}$ s$^{-1}$), the minimum energy of CR that can penetrate fully the cloud is $E_\mathrm{p}\approx2$ GeV. This is shown in Fig. \ref{fig:penetration} as the intersection of the dashed line representing the pp loss timescale and the black solid line representing the diffusion timescale in the case of slow diffusion. For faster diffusion, the minimum energy is even lower. This conclusion does not depend on the preperties of the accelerator, so as to say, it is valid both for continuous and impulsive accelerators, but depends on the diffusion coefficient, the density of the gas and the size of the cloud. Therefore, the CR background density is always added to the CR spectrum from the accelerator itself in the cases shown in Fig. \ref{fig:penetration}. From the latter figure, thus, it is possible to derive the degree of penetration to the core of the cloud for CRs of different energies, assuming a flat density distribution.

\section{CTA response}\label{sec:flat}
\subsection{Detectability}
\begin{figure}[!h]
\begin{center}
\includegraphics[width=0.6\linewidth,angle=270]{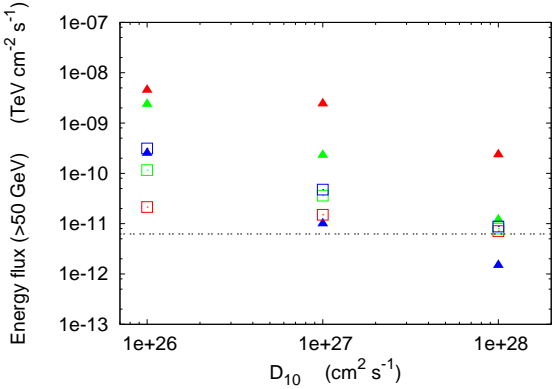}
\caption{Energy flux, for the set of permutations described in the text. Filled triangles are for impulsive cases and squares are for continuous acceleration (accelerator age of $10^{3,4,5}$ years are represented by red, green, blue markers, respectively). The dotted line represents the CTA integral sensitivity above 50 GeV as in \cite{dc2010}, scaled for 1 degree radial extension.} 
 \label{fig:fluxvsD}
\end{center}
\end{figure}

The \gr emission has been simulated for a molecular cloud with a mass of $M_\mathrm{5}=10^5 M_\odot$ and a radius of 20 pc (hence with an average density of $n_\mathrm{H}=130$ cm$^{-3}$) located at a distance of $d=$1 kpc, with the accelerator parameters given above for a continuous and impulsive source. Giant molecular clouds as close as $\sim$1 kpc distance might be uncommon, however for a list see \cite{dame_co}. The angular extension is $\sim1$ deg in radius. 
Most of the results can be rescaled to an arbitrary cloud mass and distance by recalling that, for a given CR intensity in the cloud, the expected gamma ray flux is expected to scale as $\propto M_\mathrm{5}/d^2$ and the cloud apparent size as $\propto R_\mathrm{cloud}/d$.
\gr fluxes were calculated for a permutation of age, acceleration mode, and diffusion coefficient as outlined in Section \ref{sec:accelerator}.
Fig. \ref{fig:fluxvsD} shows that for such permutations, all the cases can be detected with 50 hrs of CTA observations except for the case of an old impulsive accelerator and fast diffusion. The sensitivity of the instrument is scaled for the extension of the source, as detailed in Sec. \ref{sec:spectralfeatures}. The calculated integral fluxes are very similar for some of the permutations. 
For example, in the case of fast diffusion and continuous acceleration, the calculated flux is constant with age (red, green and blue squares at $D_{10}=10^{28}$ cm$^2$ s$^{-1}$). This can be explained by the fact that the bulk of the particles contributing to \gr emission at energies above 10 GeV can diffuse out of the cloud in a time smaller than the age of the accelerator. Therefore only the latest generation of accelerated particles contributes to the \gr signal and a steady state is reached. 

Once a spectrum and source morphology are obtained using CTA
these would act as diagnostic tools with which to
reconstruct the initial parameters of the accelerator.

\subsection{Spectral features}\label{sec:spectralfeatures}
\begin{figure}[!h]
\begin{center}
\includegraphics[width=0.49\textwidth]{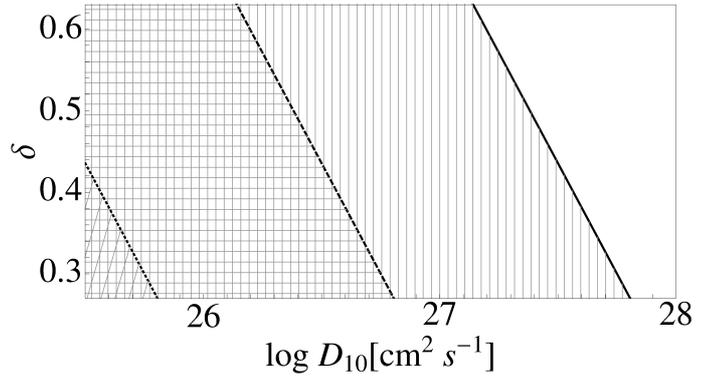}
\caption{Boundaries for a break at energies higher than 70 GeV (in the emitted \gr spectrum) for different ages of the accelerator ($10^{3,4,5}$ years are
represented by solid, dashed, dotted curves, respectively). The parameter space at the left of the boundaries results in a break in the spectrum within the energy domain of CTA.} 
 \label{fig:breakboundaries}
\end{center}
\end{figure}

A constraint on the diffusion coefficient can come from the identification of a break in the \gr spectrum integrated from the entire cloud region. The break can be related to the minimum energy that can diffuse in the entire cloud over a timescale comparable to the age of the accelerator. By equating Eq. \ref{eq:taudiff}, which gives the diffusion timescale, with the age of the accelerator, one obtains the constraints shown in Fig. \ref{fig:breakboundaries}, which correspond to:
\begin{equation}\label{eq:e_break}
 E_\mathrm{p, break}=10\left(\frac{R_\mathrm{cloud}^2}{6 D_\mathrm{10} t_\mathrm{age}}\right)^{1/\delta} \textrm{GeV}.
\end{equation}
\begin{figure}[!h]
\begin{center}
\includegraphics[width=0.98\linewidth]{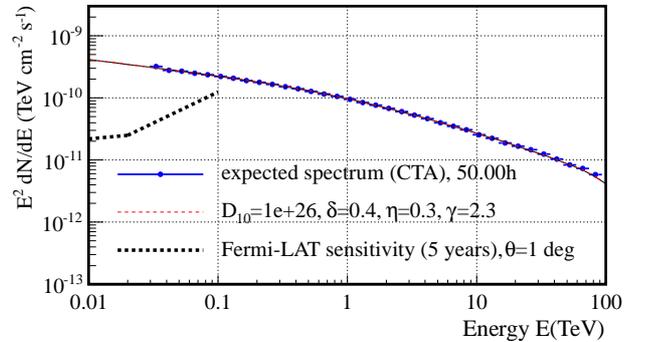}
\includegraphics[width=0.98\linewidth]{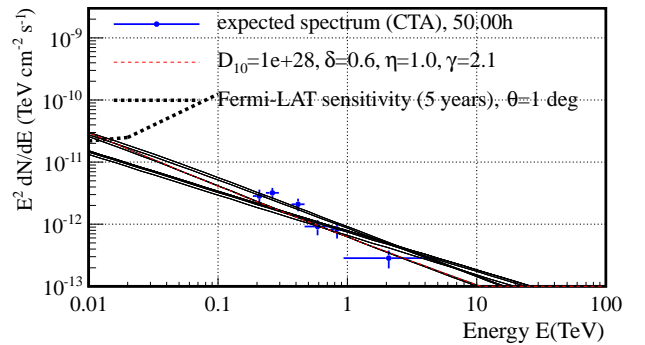}
\caption{CTA expected performances in the reconstruction of the intrinsic spectral model. Top panel: $D_{10}=10^{26}$ cm$^2$ s$^{-1}$, $\delta=0.4$, $\gamma=2.3$, $\eta=1/3$. Bottom panel: $D_{10}=10^{28}$ cm$^2$ s$^{-1}$, $\delta=0.6$, $\gamma=2.1$, $\eta=1$. The blue points illustrate one of the possible realizations of the simulated spectral points from 50 hours of CTA observation time, the red line identifies the intrinsic spectrum and the black lines represent the accepted models ($\chi^2/dof+1$). The red intrinsic spectrum is superimposed to one of the accepted models, which, for the case shown in the top panel, is superimposed to the only accepted model.}
 \label{fig:ctarespchisqdiff}
\end{center}
\end{figure}
For scenarios with fast diffusion ($D_{10}=10^{28}$ cm$^2$ s$^{-1}$) and energy dependence parameter in the range $\delta=[0.3 .. 0.6]$, the corresponding break in \gr emission will always be at energies below the CTA energy acceptance. This is accurate for impulsive accelerators. In the continuous acceleration case, new injections of high energy particles will smooth the effect of a break.
At energies higher than the break given by Eq. \ref{eq:e_break}, the particle spectrum will follow a power-law form composed of the slope of the injection spectrum and the energy dependence of the diffusion coefficient (i.e $\gamma+\delta$). Therefore the \gr emission will show a power-law behavior, thereby reducing the ability to constrain the parameter space from \gr data in the cases when $E_\mathrm{break}$ is below 70 GeV.

The reconstruction of a break is indeed a powerful tool. If a break is not present, and the CTA detected spectrum is a simple power-law, a plethora of models will fit the spectral points.
An example is given in the following, where the parameters are searched for in a grid for the intervals given in the Appendix \ref{app:grid}. 

Let us assume that a molecular cloud with measured mass and distance is detected at
TeV energies. Let us further assume that from multiwavelength observations we
identified a possible accelerator of CRs responsible for the gamma ray emission
and that an estimate of the age of that accelerator is known. In Fig. \ref{fig:ctarespchisqdiff} we show the
simulated spectra for the gamma ray emission for such an accelerator, which is
assumed to be impulsive and with an age of $10^4$ years.
The other parameters are varied on a grid (see Appendix \ref{app:grid}) and the corresponding observed spectrum is simulated from CTA ``responses''.
By using a $\chi^2$ minimization procedure, the best fitting model belonging to the parameter space grid is found, along with a
representative sample of the models within the $\chi^2/dof+1$ contour ($dof$=degrees of freedom), which is those models with a $\chi^2/dof$ not in excess of unity from the best fit model. The top panel of Fig. \ref{fig:ctarespchisqdiff} shows the case of $D_\mathrm{10}=10^{26}$ cm$^2$ s$^{-1}$, $\delta=0.4$, $\gamma=2.3$, $\eta=1/3$. Thanks to high flux reached in this case and the presence of a break, the spectrum is reconstructed easily to the intrinsic parameters, with a break in the \gr spectrum at $E\approx2$ TeV and slopes $\Gamma_\mathrm{1}=2.3$ and $\Gamma_\mathrm{2}=2.7$ below and above the break, respectively. 
However, for lower fluxes, the reconstruction can be more uncertain. This is shown in Fig. \ref{fig:ctarespchisqdiff} (bottom panel), where it can be seen that the intrinsic model ($D_{10}=10^{28}$ cm$^2$ s$^{-1}$, $\delta=0.6$, $\gamma=2.1$, $\eta=1$) does not even provide the best fit. 
\begin{figure*}[!ht]
\begin{center}
\includegraphics[width=0.49\linewidth]{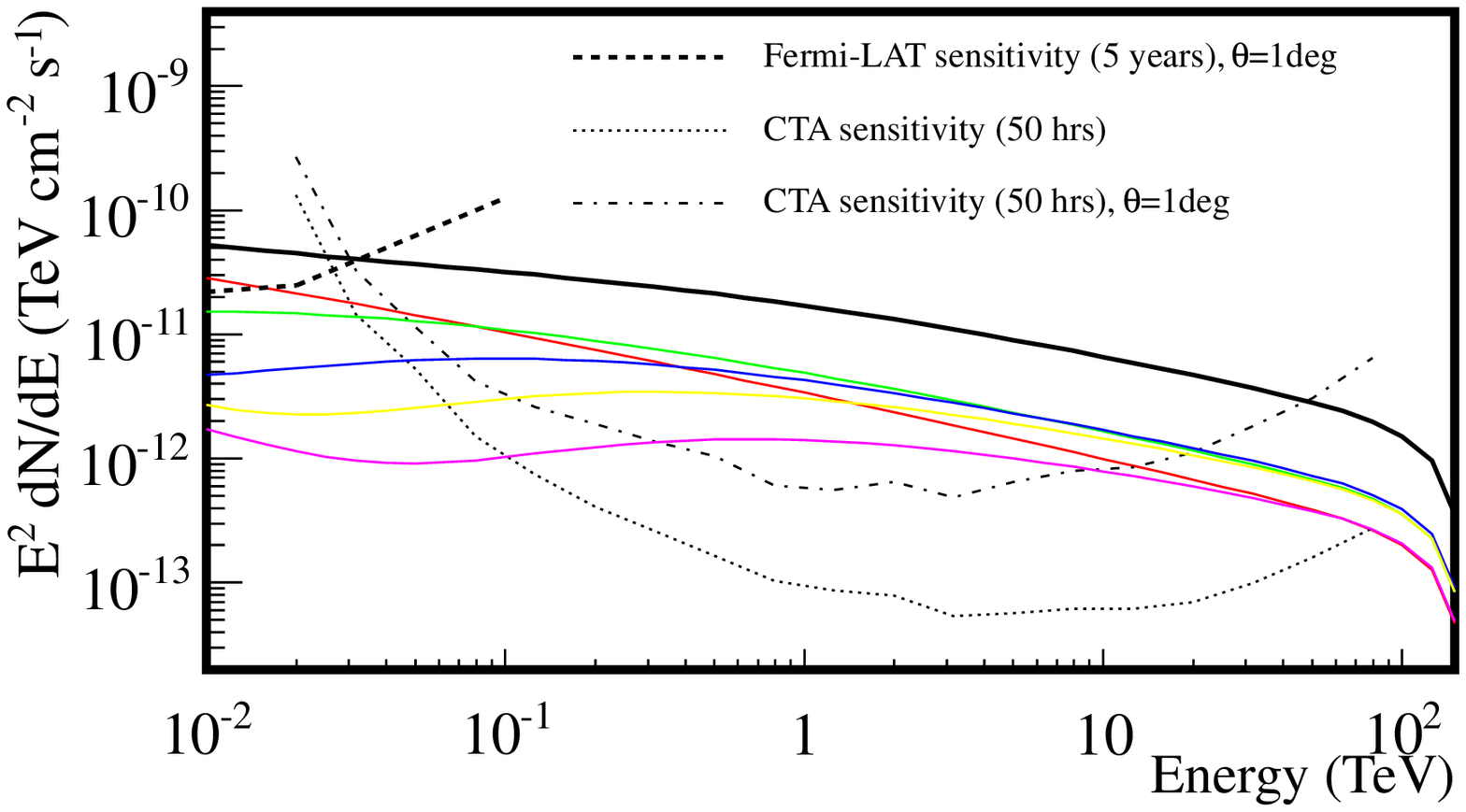}
\includegraphics[width=0.49\linewidth]{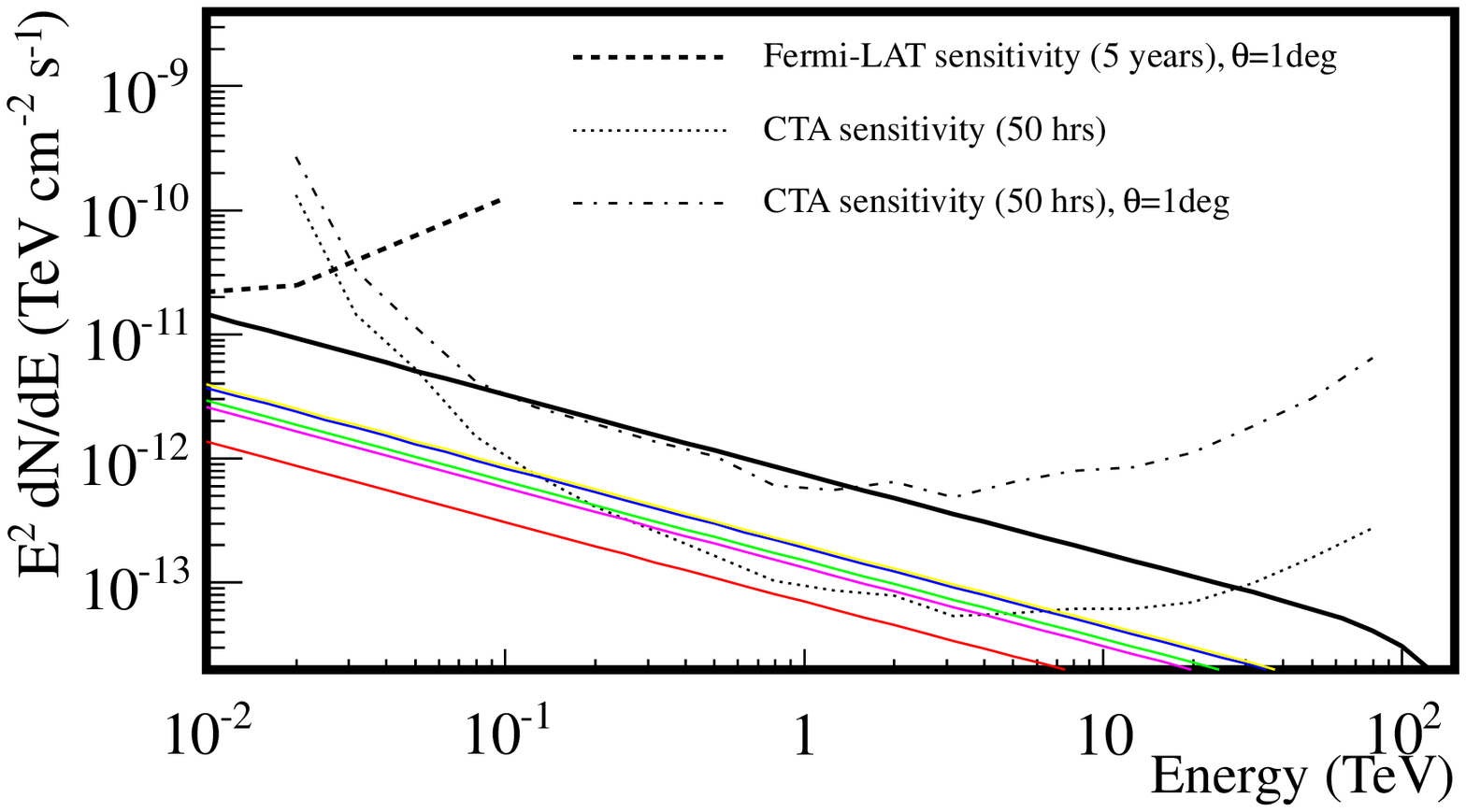}
\caption{The \gr flux of the 5 shells considered. The color code goes  from the inner to the outer shell: red, green, blue, yellow, magenta. The black line is the total flux from the molecular cloud, i.e. the sum of the 5 shells, and can be compared with Fig. 10 of \cite{aha_ato}. The two panels show the example for a continuous accelerator, with $D_\mathrm{10}=10^{26}$ cm$^2$ s$^{-1}$ (slow diffusion, left) and $D_\mathrm{10}=10^{28}$ cm$^2$ s$^{-1}$ (fast diffusion, right). In the case of fast diffusion, the spectrum is similar for all the shells, in particular the blue and yellow lines are coincident. The accelerator age is $10^4$ years.} 
 \label{fig:flux_5wedge_26_cont}
\end{center}
\end{figure*}
\begin{figure*}[!ht]
\begin{center}
\includegraphics[width=0.49\linewidth]{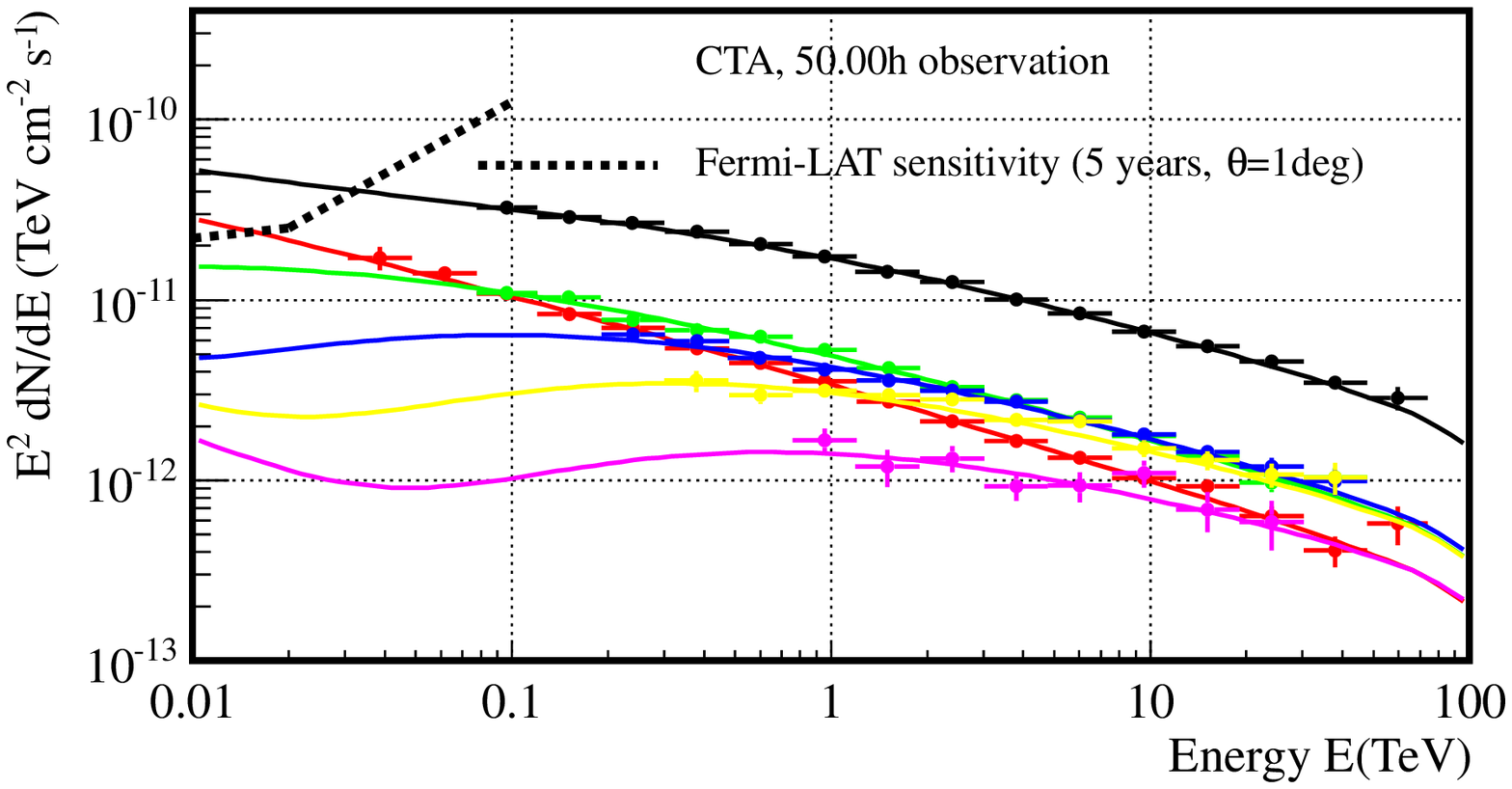}
\includegraphics[width=0.49\linewidth]{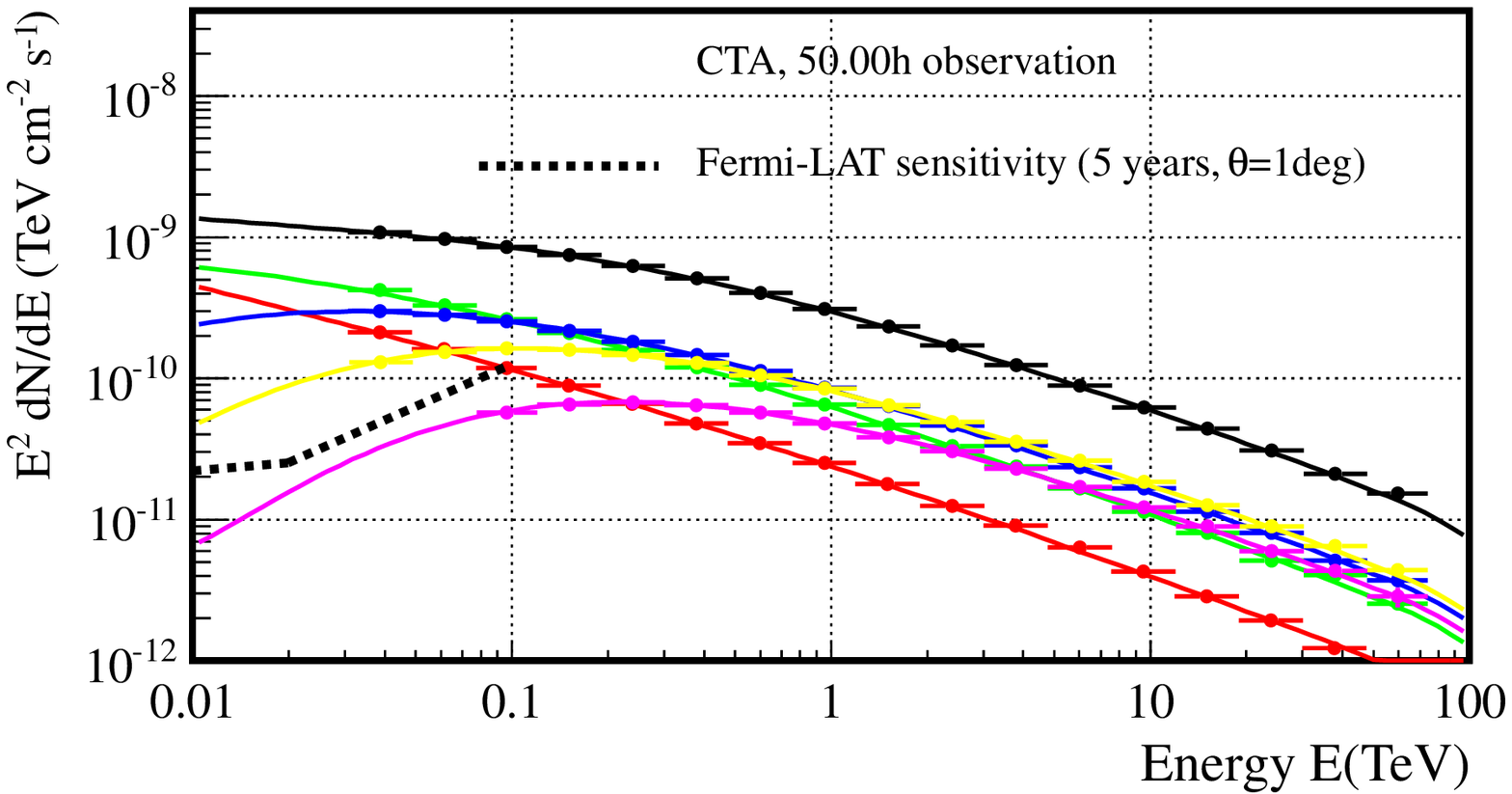}
\caption{CTA response to the scenario with $D_\mathrm{10}=10^{26}$ cm$^2$ s$^{-1}$ and an age of the accelerator of $10^{4}$ years; both for continuous acceleration (left) and impulsive acceleration (right), investigated for a 50 hours integration time. The color code follows that of Fig. \ref{fig:flux_5wedge_26_cont}.}
 \label{fig:ctaresp}
\end{center}
\end{figure*}

Additional and more stringent constraints can come from studying the spatial dependence of the \gr spectrum from the inner to the outer region, which depends on the diffusion inside the cloud.
We slice the expected emission, projected onto the sky, in concentric shells (see Appendix). We choose the linear size of the shells to match the expected angular resolution of CTA at the lowest energies resolvable by the array \citep[e.g. $\sim$0.25$^\circ$ at 50 GeV, see][]{dc2010}, with 5 concentric shells, with shell 1 being the closest to the accelerator. Fig. \ref{fig:flux_5wedge_26_cont} shows an example of the predicted fluxes and Fig. \ref{fig:ctaresp} shows the expected CTA energy spectra.

Fig. \ref{fig:flux_5wedge_26_cont} shows also the expected CTA sensitivity scaled with the size of the source.
In order to obtain the CTA sensitivity for an extended source, the point-source sensitivity is taken from \cite{dc2010} and then scaled with an appropriate energy dependent factor. This factor is related to the optimal cut on the angular size of the source.
The ratio between sensitivity for a point-like source (PS) and an extended (EXT) source is $\Theta_\mathrm{EXT}/\Theta_\mathrm{PS}$.
In Fig. \ref{fig:flux_5wedge_26_cont} we show the PS sensitivity for CTA together with the scaled one for $\Theta_\mathrm{EXT}$. We show this to illustrate the capability of CTA. However, to calculate the spectral points and profiles from CTA simulated observations, we followed the procedure detailed in \citet{apissue_mc}. To evaluate the spectral CTA response for each shell, we simulate the background counts coming from a region as extended as the outer border of each shell. Contamination from adjacent shell is not taken into account.

The LAT instrument on board of the \textit{Fermi} satellite will provide at least 5 years worth of data by the time that the full CTA array will be in operation. Figs. \ref{fig:flux_5wedge_26_cont} and \ref{fig:ctaresp} show the expected 5 year point source sensitivity for Fermi/LAT. This is calculated from the 1 year sensitivity in \citep{atwood2009}, linearly scaled with time at high energies ($>10$ GeV). The linear scaling is expected in the signal limited regime. 

The simulations show that the total cloud emission, as well as that from some individual shells, are well above the limit of detection with CTA.
For the particular case in Fig. \ref{fig:ctaresp}, the low energy particles still have not diffused out to the outer shells (the last shell is in magenta in the plot). Indeed the concave shape of the spectrum evidences the dominant contribution from the CR background at low energies on the outer shells. The concavity in the spectrum is expected at energies below the energy range of CTA. However, it will be possible to distinguish a hardening of the spectrum for shells with increasing distances from the accelerator.
These features can be seen for middle age accelerators (age=$10^4$ years) only for $D_\mathrm{10}=10^{26}$ cm$^2$ s$^{-1}$. For faster acceleration, the spectra of all shells will be a simple power-law with same photon index, thus not distinguishable. 
\subsection{Morphology}\label{sec:morphology}
\begin{figure}[!h]
\begin{center}
\includegraphics[width=0.98\linewidth]{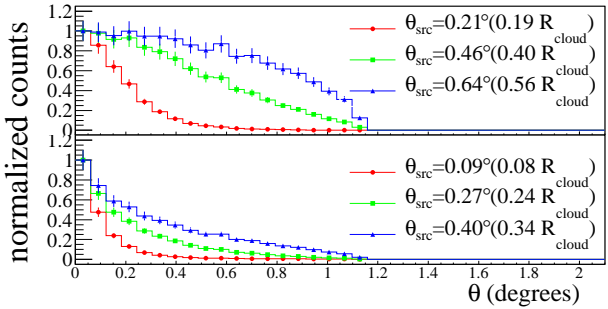}
\caption{Profiles of the photon count, normalized to the respective maximum. Here we show the example of impulsive (\textbf{Top}) and continuous (\textbf{Bottom}) acceleration, $D_{10}=10^{26}$ cm$^2$ s$^{-1}$, at an accelerator age of, from bottom to top,  $10^{3}$ (red), $10^{4}$ years (green), and $10^{5}$ years (blue). Error bars are set to 10\% of the count number, to mimic the expected error on the effective area for the array used in the counts determination.}
 \label{fig:profiles}
\end{center}
\end{figure}

The different scenarios can also be disentangled by investigating the morphology and extension of the emission region. We used the fluxes and shell area from the scenarios described above in order to simulate an excess map from the simulated response functions of CTA, as detailed in \citet{apissue_mc}.
We then create a profile from the excess map, integrating azimuthally the counts for bins in increasing angular distance from the center of the map. The profiles are then weighted by the bin area.
The shape of such profile depends on the parameters of the simulated scenario, an example of which is shown in Fig. \ref{fig:profiles}. These profiles are easily distinguishable from one another. Extensions of the \gr emission depends on the parameters studied here, with some general trends. Older sources are always more extended than younger sources, as the lower energy particles will have diffused further from the center of the cloud and thus from the accelerator. Emission due to continuous accelerators will present steeper profiles due to the freshly accelerated particles in the center of the source. Faster diffusion also leads to a larger extension. The profiles in Fig. \ref{fig:profiles} are normalized to their respective maximum counts, with flux decreasing with age in the impulsive case and opposite behavior in the continuous case. However we do not expect many of these high flux objects, therefore it is useful to predict the maximum distance at which an object is expected to be detected and resolved, depending on its intrinsic luminosity.

The maximum distance ($d_\mathrm{max}$) at which a source can be detected depends on the sensitivity of the instrument and on the angular size of the source, $\theta_\mathrm{src}=\xi R_\mathrm{cloud}/d$ ($\xi$ represents the fraction of the cloud radius that contains the 68\% of the counts in profiles as shown in Fig. \ref{fig:profiles}). 
For a source of constant luminosity, $d_\mathrm{max}^2=  L_\mathrm{src} / \left(4 \pi F_\mathrm{sens} \right)$, where $L_\mathrm{src}= 4 \pi F_\mathrm{src}  d_\mathrm{src}^2$ is the isotropic luminosity of the source. The sensitivity of the array, for an extended source, is $F_\mathrm{sens}(d)=F_\mathrm{PS} \theta_\mathrm{src} / \theta_\mathrm{PS}= F_\mathrm{PS} /\theta_\mathrm{PS} \times \xi R_\mathrm{cloud} / d$, where $F_\mathrm{PS}$ is the expected point-source sensitivity of the array and $\theta_\mathrm{PS}$ its angular resolution. For CTA, the point-source sensitivity is $F_\mathrm{PS,CTA}(>50\textrm{GeV})\approx 10^{-12} \mathrm{ TeV/cm^2/s}$ and $\theta_\mathrm{PS, CTA}\approx0.25^\circ$ its angular resolution, at low energies \citep{dc2010}. Therefore, for $F_\mathrm{sens}(d_\mathrm{max})$, we will have:
\begin{eqnarray}\label{eq:detechorizon}
 d_\mathrm{max}&\approx& 14 \left(\frac{L_\mathrm{src}}{4\times10^{33} \mathrm{erg s^{-1}}}\right) \left(\frac{F_\mathrm{PS,CTA}}{F_\mathrm{PS}}\right) \left(\frac{\theta_\mathrm{PS}}{0.25^\circ}\right) \nonumber \\
	       & \times& \left(\frac{10 \mathrm{pc}}{\xi R_\mathrm{cloud}}\right) \textrm{kpc,}
\end{eqnarray}

The maximum distance at which a source can be resolved is when the angular extension of the cloud equals the PSF of the instrument ($\theta_\mathrm{src}\equiv\theta_\mathrm{PS}$), therefore the maximum distance is
\begin{equation}
 d_\mathrm{res,max}\approx 6\left(\frac{L_\mathrm{src}}{4\times10^{33} \mathrm{erg s^{-1}}}\right)^{1/2} \left(\frac{F_\mathrm{PS,CTA}}{F_\mathrm{PS}}\right)^{1/2} \textrm{kpc.}
\end{equation}

\begin{figure}[!h]
\begin{center}
\includegraphics[width=0.98\linewidth]{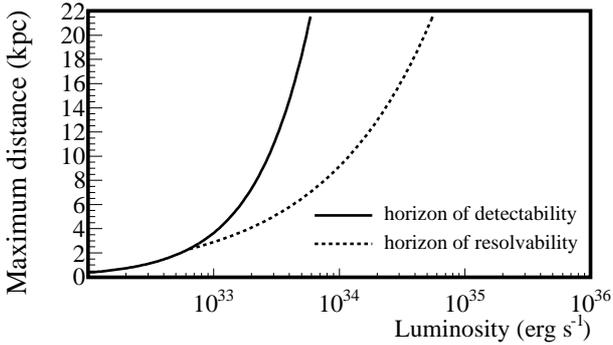}
\caption{Horizons of detectability (solid) and resolvability (dotted) as a function of isotropic luminosity ($\xi=0.5$, other parameters as in text).} 
 \label{fig:horizons}
\end{center}
\end{figure}

At this distance we will be able to discriminate between the point-source and extended case, but the exact determination of the extension might need longer observation times.
Fig. \ref{fig:horizons} shows the evolution of the maximum distances of detection and resolvability as a function of the luminosity of the source. 
As an example, let us consider a passive cloud, i.e. a case where only the CR background is included. With this assumption, the expected signal is \citep{aha_passive,gabici_review}:
\begin{eqnarray}\label{eq:passivecloud}
&& F(> E_\gamma) \sim  
  1 \times 10^{-13}  \kappa \left(\frac{E}{\textrm{TeV}}\right)^{-1.7}  \nonumber \\
&& \hspace{2cm} \times \left(\frac{M}{10^5 M_\odot}\right) \left(\frac{D}{1\textrm{kpc}}\right)^{-2} \mathrm{cm}^{-2} \mathrm{s}^{-1},  
\end{eqnarray}
where $\kappa$ is the enhancement factor of CRs, assumed to be unity for passive clouds. Therefore a cloud of $10^5 M_\odot$ would be detected out to only $\sim 1$ kpc, due to its expected isotropic luminosity of $L_\mathrm{iso} \approx 2 \times 10^{32} \textrm{ erg s}\mathrm{^{-1}}$. If such cloud was to be more compact than the ones investigated here, the horizon of its detectability would be larger, according to the scaling given in Eq. \ref{eq:detechorizon}.
It has to be noted that enhancing the CR content of the cloud by a factor $\kappa>20$ would allow the detection of a cloud of such mass in the entire galaxy. For comparison, $\kappa\sim20$ would correspond to one of the cases exemplified in Fig. \ref{fig:fluxvsD}, specifically a continuous accelerator of $10^4$ year age and $D_{10}=10^{27}$ cm$^2$ s$^{-1}$.

\section{Passive clouds: giants and cloudlets}
\begin{figure}[!h]
\begin{center}
\includegraphics[width=0.9 \linewidth]{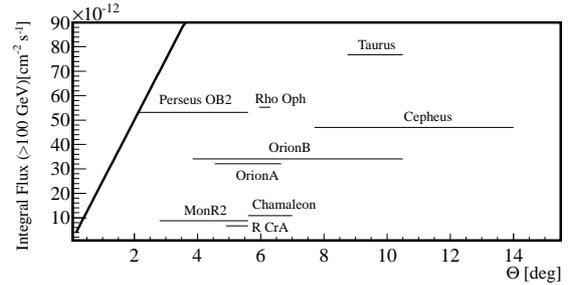}
\caption{Expected flux from the GMC in the Gould Belt vs their extension \citep[extension is 70\% of the boundaries listed in ][]{dame_co}. Expected flux ($>100$ GeV) is calculated through Eq. \ref{eq:passivecloud}, with $\kappa=1$. The solid line is the on-axis sensitivity of CTA \citep[$>100$ GeV, 50 hours;][]{dc2010}, linearly scaled with the extension of the source. This optimistic scaling holds for an infinite field of view with flat acceptance, while we expect a degradation of sensitivity with increasing angular distance from the center of the field of view.}
 \label{fig:gouldbelt}
\end{center}
\end{figure}

Giant molecular clouds (GMC) as close as $\sim$1 kpc distance are uncommon \citep[for a list see][]{dame_co}.
These are passive clouds (i.e. only the CR background is included) that would also be interesting for detection. The feasibility of a detection depends mainly on their mass, distance and most of all extension of the gamma-ray emission, as already discussed in \cite{gabici_cta}.
Here we assume that the extension is $\sim$0.7 of the boundaries listed in \cite{dame_co} together with  the mass and distances given in that paper. The assumption on the extension comes from the fact that we need to consider the 68\% containment radius of the \gr emission. We estimate the \gr flux from Eq. \ref{eq:passivecloud}, with $\kappa=1$, and we compare it to the integral sensitivity given in \citep[$>100$ GeV, 50 hours;][]{dc2010} and its scaling for the extension of the source (i.e. $\propto \theta_\mathrm{src}/\theta_\mathrm{PS}$, where $\theta_\mathrm{PS}$=0.1). The scaling of the sensitivity given here is valid only for an infinite field of view with flat acceptance, while we expect a degradation of sensitivity with increasing angular distance from the center of the field of view \citep[for a discussion see][]{survey_cta}. Many of the clouds are very extended with respect to the expected angular acceptance of CTA and would need either to be scanned or to be observed in divergent mode, therefore increasing the observation time for a mapping of the entire object \citep[for a general survey in divergent mode see][]{survey_cta}. Expected fluxes and the corresponding scaled sensitivity are given in Fig. \ref{fig:gouldbelt}. The ranges in extension refer to the cloud boundaries in latitude and longitude.

The prospects for detection with CTA are slim due to the very large extension of the clouds.
However, there is an abundance of smaller clouds at closer distances, that have been already detected at HE $\gamma$-rays. Following the averages derived from the cloudlet population studied in \cite{torres2005}, we simulate a 40 M$_\odot$ cloud at a distance of 150 pc and an extension of 2.8 pc (1.08$^\circ$ in angular size). The resulting density of the cloudlet is similar to that of the larger cloud studied in the paragraph above, $n_\mathrm{H}=140$ cm$^{-3}$. Therefore we expect total penetration of CR sea for energies above $E_p > 1$ GeV for such small radial extensions.
However, it is unlikely to have such cloudlets hosting a powerful accelerator. So we consider them also as passive clouds. With this assumption, the expected signal is given by Eq. \ref{eq:passivecloud}, with $\kappa=1$. This translates for the cloudlet considered into $F(> 1 \textrm{TeV}) \sim 2 \times 10^{-15} \mathrm{cm}^{-2} \mathrm{s}^{-1} \sim 0.1\permil \,$ CU, where CU stands for Crab Units and represents the integral flux of the Crab nebula above 1 TeV.
This flux would not be easily detectable (the expected CTA sensitivity is of $\sim 1\permil \, CU$). Nonetheless, a stacking approach on a population of roughly 100 clouds --a list of which is used by \cite{torres2005}-- would lead to a detection also in VHE $\gamma$-rays, implying a good test for the isotropy of the CR sea in the local neighborhood. 
\section{Peaked density profile of the molecular cloud}
In this section we describe qualitatively the expectation for a different density distribution of the target material inside the cloud. Indeed, the density of the molecular cloud might not be constant throughout its extension. 
Therefore we study the case of a cloud with a density profile of the form
\begin{eqnarray}
 n_\mathrm{H}(r) & = & \frac{n_0}{\left(1+{r} / {R_\mathrm{C}}\right)^\alphan},
  \label{param}
\end{eqnarray}
where ${R_\mathrm{C}}$ is the core radius and $n_0$ is the density at the center of the cloud.
Following the observations by \citet{crutcher} we assume that the cloud magnetic field scales with density as
\begin{eqnarray}
 B(r) & \sim & 100 \left(\frac{n_\mathrm{H}(r)}{10^{4}\mathrm{cm}^{-3}}\right)^{1/2}.
\end{eqnarray}
This will in turn affect the value of the CR diffusion coefficient. We parametrize
the effect of the magnetic field as:
\begin{eqnarray}
 D\left(E,r\right) & = & D_0\left(\frac{E/\mathrm{GeV}}{B(r)/3\mu\mathrm{G}}\right)^\delta.
\end{eqnarray}
The mass of the interacting target material can be calculated following
\begin{eqnarray}
 M &=& \int_{R_1}^{R_2}4\pi m_\mathrm{p}r^2n_\mathrm{H}(r)dr \nonumber \\
   &=& 4\pi m_\mathrm{p}n_0 R_\mathrm{C}^3\left[\frac{p^{3-\alphan}}{3-\alphan}-\frac{2 p^{2-\alphan}}{2-\alphan}+\frac{p^{1-\alphan}}{1-\alphan}\right]_{R_1}^{R_2},
\end{eqnarray}
where $p=(1+r/R_\mathrm{C})$. The case of $\alphan=0$ reduces to a flat density profile investigated in Section \ref{sec:flat}. 
The total mass $M$ is obtained with $R_1=0$ and $R_2=R_\mathrm{cloud}$. If the mass of the cloud is known, the central density can be derived from the formula above.

We again assume $M=10^5 M_\odot$ and $R_\mathrm{max}=$20 pc. From this assumption we can calculate the central density depending on the $\alphan$ chosen (with the core radius $R_\mathrm{C}$=0.5 pc {fixed}). This allows us to calculate for each radius, $n_\mathrm{H},B,D$ from Eq. (\ref{param}). 

\begin{figure}[!h]
\begin{center}
\includegraphics[width=0.6 \linewidth, angle=270]{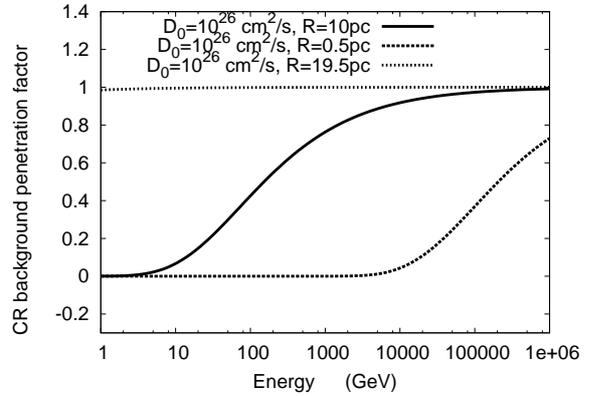}
\caption{Cosmic-ray penetration factor, $e^{-\tau_*(E,r)}$, for $\alphan$=2 as a function of the particle energy for the center (R=0.5pc, dashed), the middle (R=10pc, solid), and the border of the cloud (R=19.5pc dotted), in the case of $D_{10}=10^{26}$ cm$^2$ s$^{-1}$.}
 \label{fig:penfactor}
\end{center}
\end{figure}

With increasing densities, and correspondingly stronger magnetic fields, diffusion becomes much slower and the timescales for pp losses faster, resulting in a slower penetration of the CR sea. Because of this, the penetration of the CR sea into the cloud is not complete, and we cannot any longer assume that the sea is constant throughout the cloud. To compute the level of penetration of the CR sea into the cloud, we follow here the approach by \citet{prot08} and treat the penetration as an analog of optical thickness, where the CR intensity $I_\textrm{CR}=e^{-\tau_*(E,r)} I_\textrm{CR}(E,r)$ and $\tau_*(E,r)=\tau_\mathrm{a}\left(\tau_\mathrm{a}+\tau_\mathrm{s}\right)$ can be decomposed in two terms, analogs of absorption and scattering:
\begin{eqnarray}\label{eq:depth_abs}
\tau_\mathrm{a}&\approx&\int_r^R 2 \kappa n_\mathrm{H}(r)\sigma_\mathrm{pp}dr\nonumber \\
               &=&2 \kappa \sigma_\mathrm{pp} n_0 R_\mathrm{C} \left(1-\alphan\right)^{-1}\ \left[p^{1-\alphan}\right]_{r}^{R},
\end{eqnarray}
\begin{eqnarray}
\tau_\mathrm{s}&\approx&\int_r^R \frac{c}{3 D(E,r)} dr \nonumber\\
               &=&\frac{c R_\mathrm{C}}{3 D_0 B_0 E_\mathrm{p}^\delta}\left(1-\frac{\alphan \delta}{2}\right)^{-1} \left[p^{1-\frac{\alphan \delta}{2}}\right]_{r}^{R}.
\end{eqnarray}
Examples of CR penetration as a function of energy are shown in Fig. \ref{fig:penfactor}. Comparing the penetration factors and density profiles, we will expect a dominant contribution from the accelerator in the inner shells. The low number density of target material in the outer shell and the correspondent higher diffusion coefficient assure full penetration of the CR background.  On the other hand, only a fraction of the accelerated
particles will manage to diffuse out to the outermost shells. Indeed,
because of the higher densities found in the core of the cloud, the
dominant emission in the outer shells will come from the CR
background.
With these assumptions, we expect to detect a source with a steep surface brightness profile.

\section{Concluding remarks}

CTA is the forthcoming array of IACTs, and one of its most important physics goals is the study of the origin and propagation of CR. 
We have investigated its capabilities for the study of CR diffusion in molecular clouds, by making a phase space exploration of different cases which could be observable with such facility. This complements the study presented by \cite{gabici_cta}.
We have showed theoretical predictions for VHE \grs fluxes from an accelerator inside a massive cloud. The simulations using the CTA response for 50h observation time indicate that it will be possible to constrain the diffusion coefficient parameter space.
To have the accelerator inside the cloud is an idealized case. However, it allows us to study the impact of CTA observations on expected spectral and morphological features due to an overdensity of CR in the cloud over the CR sea.

We expect to be able to detect \gr emission from massive molecular clouds with CTA observations. Specifically, for clouds of $10^5 M_\odot$ and $\theta_\mathrm{src}\sim 0.5^\circ$, we expect detection even in the case of passive cloud, i.e. only permeated by the CR background, out to distances of $\lesssim 1$ kpc. Clouds with the same mass, but with an enhanced population of CR ($\kappa>20$) could be detected throughout the galaxy.
GMC in the Gould Belt are very massive but very extended objects, therefore rendering problematic the prospect of detection for these clouds.

Spectral features will aid derivation of constraints, especially the possible identification of breaks. Moreover, emission from extended objects could be divided in subregions. The superior angular resolution and sensitivity of CTA, with respect to the current generation of instruments, will permit the study of spatial spectral evolution with increasing distance
from the accelerator, which retains the footprint of
the underlying particle distribution and its diffusion.
Regarding the source profiles, older sources are expected to have larger extension. A faster diffusion coefficient will have the same effect. It is expected that, if the observed molecular cloud exhibits a peaked density profile, the emission profile will also be peaked. Indeed the outer shell of the molecular cloud might fall below detection with CTA sensitivity, and thus prevent the detection of the full extension of the cloud. Therefore the knowledge of the properties of molecular clouds is important to aid the study of the objects described here. Information on the distribution of the target material will be of great help in order to reconstruct the parameters relative to the acceleration and diffusion of charged particles in the cloud in those cases.

At the beginning of operation, CTA might be used to survey the galactic plane. Two types of tests could be performed, even with small integration times for a given position in the sky. The observation of a typical impulsive source like a SNR hosted in a molecular cloud with similar mass and distance as the ones investigated here, even if the age of the accelerator is of the order of $10^5$ years, will constrain the diffusion coefficient (provided a precise knowledge of its distance), with detection assured for slow diffusion even with a few hours integration. While a source with the high fluxes investigated here might be rare (and would probably be already detected in TeV even though with much less detail in the spectral reconstruction), we have shown in Section \ref{sec:morphology} that we will be able to probe a significant fraction of the galaxy for even lower fluxes. Therefore, at the completion of the galactic plane survey, we will be able to study a numerous population, thus contributing to the study of the properties of accelerators and the propagation mode of CRs in molecular cloud.
\begin{appendix}
\section{Grid construction}\label{app:grid}
The sphere of emission has been divided in a regular cartesian grid (\textbf{x},\textbf{y},\textbf{z}). 
The approximation is checked with the calculation of the total mass, correctly reproduced. Also the \cite{aha_ato} curves are reasonably reproduced by our code. 
Throught the paper the parameter space investigated is limited in a grid formed by the following parameters:
\begin{itemize}
 \item Diffusion coefficient: Slow to fast (e.g., $D_{10}=10^{26}, 10^{27}, 10^{28}$ cm$^{2}$ s$^{-1}$);
\item Diffusion coefficient energy dependence: $\delta= 0.3, 0.4, 0.5, 0.6]$;	
\item  Age of the accelerator: $10^3, 10^4, 10^5$ years
\item  Type of accelerator: Impulsive / Continuous
 \item  Spectrum of injection: $\gamma$= $2.0, 2.1, 2.2, 2.3, 2.4, 2.5$;	
  \item  Fraction of energy in input ($\mathrm{W_p}=\eta10^{50}$ erg): $\eta=0.3, 1, 3$. 	
\end{itemize}
The corresponding flux is calculated for each grid block from the corresponding proton spectrum solving the diffusion equation as in \citep{aha_ato}.
The emission is then integrated along the line of sight. The integration is on the variable \textbf{y}, where a condition is posed. The maximum accepted is $y_\mathrm{max}=\sqrt{R_\mathrm{cloud}^2-z^2-x^2}$. Only a quarter of the sphere is calculated, because of isotropy.

The emission integrated along the line of sight can then be used to simulate the flux of the chosen projected shells. Throughout the paper we divide the projected emission region in five concentric shells.
The projected mass of each of the five shells is shown in Table \ref{tbl:shells}.
\begin{table}[!htb]
\centering
\caption{Characteristics of the shells}
\begin{tabular}[t]{cccl}
\hline
$R_\mathrm{in}$    &     $R_\mathrm{out}$      &    size   &  Mass$_\mathrm{proj} $   \\ 
 (pc)   &     (pc)      &    (deg)   &   $(\Msol)$   \\ 
\hline
0&4&0.23&6655\\
4&8&0.23&18590\\
8&12&0.23&27736\\
12&16&0.23&32571\\
16&20&0.23&23265\\
\hline
\end{tabular}
\label{tbl:shells}
\end{table}

\end{appendix}

\acknowledgements
We acknowledge discussions with Ana Y. Rodriguez-Marrero and many colleagues of the CTA collaboration, especially Sabrina Casanova, Elsa de Cea del Pozo, Daniela Hadasch, Jim Hinton, Konrad Bernloehr, and Abelardo Moralejo.
We acknowledge support from the Ministry of Science and the Generalitat de Catalunya, through 
the grants AYA2009-07391 and SGR2009-811, as well as by
ASPERA-EU through grant EUI-2009-04072.

\end{document}